\documentclass[aps,floatfix,pra,superscriptaddress,reprint,showpacs,10pt]{revtex4-1}
\usepackage[utf8]{inputenc}
\usepackage[dvips]{graphicx}
\usepackage{float}
\usepackage{amssymb}
\usepackage{amsmath}  
\usepackage{dsfont}
\usepackage{color}
\usepackage{array}
\usepackage[locale=US]{siunitx}	
\usepackage{bm}
\usepackage{mathrsfs}
\usepackage{pifont}
\usepackage[dvips,pdftitle={Quantum simulation of the Schwinger model},pdfauthor={Stefan Kuehn},bookmarks,linkcolor=black,citecolor=black,menucolor=black,urlcolor=black,hyperfootnotes=false,plainpages=false,pdfpagelabels,hypertexnames=false]{hyperref}

\newcommand{\su}{\ensuremath{|\mathbf{1}\rangle}}
\newcommand{\sd}{\ensuremath{|\mathbf{0}\rangle}}

\newcommand{\fz}{\ensuremath{|0\rangle}}

\newcommand{\Zn}{\ensuremath{\mathds{Z}_d}}
\newcommand{\X}{\ding{53}}

\begin{document}
\title{Quantum simulation of the Schwinger model: A study of feasibility}
\author{Stefan K\"{u}hn}
\author{J. Ignacio Cirac}
\author{Mari-Carmen Ba\~{n}uls}
\affiliation{Max-Planck-Institut f\"{u}r Quantenoptik, Hans-Kopfermann-Straße 1, 85748 Garching, Germany}
\date{\today}
\begin{abstract}
We analyze some crucial questions regarding the practical feasibility of quantum simulation for lattice gauge models. Our analysis focuses on two models suitable for the quantum simulation of the Schwinger Hamiltonian, or QED in 1+1 dimensions, which we investigate numerically using tensor networks. In particular, we explore the effect of representing the gauge degrees of freedom with finite-dimensional systems and show that the results converge rapidly; thus even with small dimensions it is possible to obtain a reasonable accuracy. We also discuss the time scales required for the adiabatic preparation of the interacting vacuum state and observe that for a suitable ramping of the interaction the required time is almost insensitive to the system size and the dimension of the physical systems. Finally, we address the possible presence of noninvariant terms in the Hamiltonian that is realized in the experiment and show that for low levels of noise it is still possible to achieve a good precision for some ground-state observables, even if the gauge symmetry is not exact in the implemented model. 
\end{abstract}
\pacs{03.67.Ac, 03.75.Be, 42.50.Ex}
\maketitle
\section{Introduction}
Gauge theories are a central part of our understanding of modern physics. A standard tool for exploring them in the nonperturbative regime is Wilson's lattice gauge theory (LGT) \cite{Wilson1974}, where the continuous theory is formulated on a discrete space-time lattice. In the context of LGT, advanced Monte Carlo simulations have been developed and it is possible to study phase diagrams \cite{Fukushima2011}, mass spectra \cite{Durr2008}, and other phenomena. However, despite the great success of these techniques there are still many problems which cannot be addressed with them, e.g., out of equilibrium dynamics or regions of the phase diagram where Monte Carlo simulations suffer from the sign problem \cite{Troyer2005,Fukushima2011}. Therefore it would be highly desirable to have new tools \cite{Trotzky2012} which overcome these problems. Quantum simulation may offer one such alternative route to tackle gauge theories, and indeed, during recent years there have been several proposals for (analog)  quantum simulators using atomic systems  \cite{Cirac2010,Zohar2011,Zohar2012,Banerjee2012,Hauke2013,Banerjee2013,Zohar2013,Zohar2013a,Tagliacozzo2013a,Tagliacozzo2013,Kapit2011}.

Quantum simulation of LGT presents a number of particular features. The quantum systems that can be controlled and manipulated to realize a quantum simulator have finite-dimensional Hilbert spaces. This is in contrast to the infinite-dimensional degrees of freedom required by continuous gauge symmetries. Therefore the models that can be realized in the experiments often correspond to finite-dimensional or truncated versions of the original gauge groups, and it is in the limit in which the dimensions get large  that the original models are recovered.

Furthermore, in quantum simulation proposals the Hamiltonian often arises as a (low-energy) effective model that governs the dynamics of atoms trapped in an optical lattice. However, in this limit the terms generated by the atomic interaction do not necessarily exhibit the gauge symmetry of the model to be simulated. In such cases, the Gauss law is imposed via a penalty term that penalizes  nonphysical states \cite{Zohar2011,Banerjee2012,Zohar2012,Hauke2013}. In particular proposals the right symmetry can be ensured  via a more fundamental conservation law \cite{Zohar2013,Zohar2013a,Banerjee2013}, however, even in these cases the presence of noise in the simulation is likely to break gauge invariance. Hence, it is a crucial question whether the nonfundamental character of the gauge symmetry in a quantum simulator will affect its expected performance.

Additionally, in order to assess the feasibility of such a simulation of LGT, the mere physical requirements also need to be analyzed. This includes scaling of resources, such as the minimal system size that will allow the observation of relevant phenomena, the time required for adiabatic preparation schemes, and the necessary noise control.

In this paper we address some of these issues by studying two possible realizations of the Schwinger model that might be suitable for ultracold atoms. We focus on proposals which have a built-in gauge symmetry, but where the gauge degrees of freedom are represented by a Hilbert space with small dimension. For these models we numerically address three questions using matrix product states (MPS) \footnote{Another direction which was recently explored is the application of MPS and more general tensor networks directly to lattice gauge theories \cite{Byrnes2002,Banuls2013,Banuls2013a,Buyens2013,Rico2013,Silvi2014,Tagliacozzo2014}} with open boundary conditions to reflect a possible experimental realization. First, we investigate how the truncation of the gauge degrees of freedom to a finite-dimensional Hilbert space affects the nature of the ground state and reveals that even a small dimension allows quite accurate predictions for the ground-state energy. Second, we examine the resources needed for adiabatic preparation of the ground state. We give evidence that the first part of the adiabatic evolution is crucial, and for the systems we study, with up to a hundred sites, the total time required for a successful preparation is practically independent of the system size. Our results also show that the Hilbert space dimension of the gauge degrees of freedom hardly affects the success of the preparation procedure. Third, we analyze the effect of imperfect gauge symmetry by studying the adiabatic preparation in the presence of noninvariant noise terms, as these might occur in an experimental realization. We quantify the level of noise up to which the results for the ground-state energy are still reasonably close to those for the noiseless case.

The rest of the paper is organized as follows. In Sec. \ref{sec:preliminaries} we briefly review the Schwinger model and explain the two particular discrete versions studied here. Furthermore, we give a brief description of the numerical methods we use. In Sec. \ref{sec:finite_d} we present our results on how the finite-dimensional Hilbert space for the gauge degrees of freedom affects the ground state. Subsequently we examine one possible scenario for the adiabatic preparation of the ground state in Sec. \ref{sec:evo} and study the effect of gauge invariance breaking noise during this procedure in Sec. \ref{sec:noisy_evo}. Finally, we conclude in Sec. \ref{sec:conclusion}.

\section{\label{sec:preliminaries}Models and methods}
The Schwinger model, or QED in $1+1$ dimensions, was introduced in \cite{Schwinger1962}. It is possibly the simplest gauge theory with matter and its compact lattice formulation shows non-trivial phenomena, like confinement, that are also observed in QCD. Therefore it is an ideal benchmark model for LGT techniques. 

A possible discrete version of the Schwinger Hamiltonian on a lattice with spacing $a$ is given by the Kogut-Susskind formulation \cite{Banks1976},
\begin{align}
\begin{aligned}
 H=&\frac{g^2a}{2}\sum_n\left(L^z_n\right)^2+m\sum_n(-1)^n\phi^\dagger_n\phi_n\\
  &-\frac{i}{2a}\sum_n\bigl(\phi^\dagger_nL^+_n\phi_{n+1}-\text{h.c}\bigr),
\end{aligned}
\label{hamiltonian}
\end{align}
where $g$ denotes the coupling constant and $m$ the fermion mass. The field $\phi_n$ is a single-component fermionic field sitting on site $n$ and the operators $L^+_n=\exp\bigl(i\theta_n\bigr)$, $L^z_n$ act on the links in between sites $n$ and $n+1$. The operators $\theta_n$ and $L^z_n$ fulfill the commutation relation $[\theta_n,L^z_m]=i\delta_{n,m}$, where $\theta_n$ can take values in $[0,2\pi]$. $L^+_n$ therefore acts as a rising operator for the electric flux on link $n$ and $L^z_n$ gives the quantized flux on the link. The physical states satisfy $G_n|\Psi\rangle=0$ $\forall n$ \cite{Kogut1975}, where  
\begin{align}
G_n=L^z_n-L^z_{n-1}-\phi^\dagger_n\phi_n+\frac{1}{2}\bigl[\mathds{1}-(-\mathds{1})^n\bigr]
\label{gausslaw}
\end{align}
are the Gauss law generators.

There are several proposals to quantum simulate the Schwinger model \cite{Banerjee2012,Banerjee2013,Zohar2013,Hauke2013}. Since the dimensions of quantum systems available for quantum simulation are finite, most proposals focus on models with finite-dimensional variables on the links that recover Hamiltonian (\ref{hamiltonian}) in the limit $d\to\infty$. One way is to simulate a quantum link model, in which the gauge variables are represented by finite-dimensional quantum spins \cite{Orland1990,Chandrasekharan1997}; another is to truncate the dimension of the link variables \footnote{A similar truncation of the Hilbert space dimension for the gauge degree of freedoms is, e.g., used in some tensor network simulations of LGT \cite{Byrnes2002,Buyens2013}}. These approaches can lead to a Hamiltonian with a gauge symmetry which is different from that of the Schwinger model.

Here we consider two particular models, one which has the same gauge symmetry as the Schwinger model despite the finite-dimensional links, and one which has a different gauge symmetry due to the finite dimension.

\subsection{Truncated compact QED (cQED) model}
The first model we examine corresponds to the proposal for the simulation of the cQED from Ref. \cite{Zohar2013}, using fermionic and bosonic atoms trapped in an optical superlattice. The fermions are sitting in the minima of one lattice forming the sites. The links are populated by an (even) number of particles $N_0=a^\dagger_na_n+b^\dagger_nb_n$, consisting of two bosonic species $A$ and $B$, sitting between the fermions in the minima of another lattice. The operators $a_n$ and $b_n$ ($a_n^\dagger$ and $b_n^\dagger$) are the annihilation (creation) operators for species $A$ and $B$ on link $n$, fulfilling the usual commutation relations. This model gives rise to a Hamiltonian of the form of (\ref{hamiltonian}) with link operators
\begin{align}
L^+_n = i\frac{a_n^\dagger b_n}{\sqrt{l(l+1)}},\quad\quad L^z_n=\frac{1}{2}(a_n^\dagger a_n-b_n^\dagger b_n),
\label{cqed_op}
\end{align}
where $l=N_0/2$, so that the link operators are angular momentum operators in the Schwinger representation. As $a^\dagger_na_n+b^\dagger_nb_n$ is a constant of motion, the number of particles on a link, $N_0$, is conserved. The dimension of the Hilbert space for each link is given by $d=N_0+1$, and in the limit $N_0\to\infty$ the link operators become pure phases that coincide with those from the Kogut-Susskind Hamiltonian. In this realization, the angular momentum conservation in the scattering between fermionic and bosonic species ensures the Gauss law, which does not have to be imposed effectively via a penalty term. 

The Hamiltonian in this case is invariant under local transformations that affect the annihilation operator for one fermion on site $n$ and its adjacent bosons as
\begin{align*}
\phi_n &\to e^{i\alpha_n}\phi_n,\\
b_{n-1}&\to e^{i\alpha_n}b_{n-1},\\
a_{n}&\to e^{-i\alpha_n}a_{n}
\end{align*}
while the operators acting on other sites and links are unchanged. The model then has the same $U(1)$ symmetry as the untruncated Schwinger model and we refer to it as the \emph{truncated cQED model}. The Hamiltonian of this model commutes with the Gauss law generators
\begin{align*}
G^\mathrm{cQED}_n = L^z_n-L^z_{n-1}-\phi^\dagger_n\phi_n+\frac{1}{2}\bigl[\mathds{1}-(-\mathds{1})^n\bigr],
\end{align*}
where the $L^z_n$-operators are given by Eq. (\ref{cqed_op}).

\subsection{$\Zn$ model}
Another possibility to represent the links with finite-dimensional objects is to substitute the infinite-dimensional $U(1)$ gauge operators in (\ref{hamiltonian}) with $\Zn$ operators. This can be realized with the link operators
\begin{align}
 L^+_n=\sum_{k=-J}^J |\varphi^{k+1}_n\rangle\langle\varphi^{k}_n|,\quad L^z_n=\sum_{k=-J}^J k|\varphi^k_n\rangle\langle\varphi^k_n|,
\label{zn_op}
\end{align}
where one needs to identify $|\varphi^{J+1}_n\rangle $ with $|\varphi^{-J}_n\rangle$. Consequently the dimension of the Hilbert space of a link is given by $d=2J+1$. As shown in Ref. \cite{DeLaTorre1998}, in the limit $d\to\infty$ these operators approach the link operators of the Kogut-Susskind Hamiltonian. 

The resulting Hamiltonian is invariant under local transformations of the fermions and adjacent links \footnote{For simplicity we show here the effect of the transformation on the basis states for the links and not the operators. One should also note that one has the freedom to add arbitrary constant phase factors to the transformation for the basis states.} as
\begin{align*}
\phi_n&\to e^{i\alpha_n}\phi_n,\\
|\varphi^k_{n-1}\rangle &\to e^{-i k\alpha_n}|\varphi^k_{n-1}\rangle,\\
|\varphi^k_n\rangle &\to e^{i k\alpha_n}|\varphi^k_n\rangle,
\end{align*}
with $\alpha_n=2\pi q/d$, $q\in\mathds{Z}$. Differently from the truncated cQED case, here only discrete phase transformations leave the Hamiltonian invariant \footnote{We call the model presented here the $\Zn$ model because of this discrete symmetry. However, one should note that it does not correspond to a $\Zn$ lattice gauge theory \cite{Horn1979}, as we use a different kinetic term for the gauge field in the Hamiltonian.}. Correspondingly the Gauss law is only fulfilled modulo $d$ and the operators that commute with the Hamiltonian are actually
\begin{align}
U^{\Zn}_n=e^{i\frac{2\pi}{d}\left(L^z_n-L^z_{n-1}-\phi^\dagger_n\phi_n+\frac{1}{2}\left[\mathds{1}-(-\mathds{1})^n\right]\right)},
\label{GL_Zn}
\end{align}
where the $L^z_n$-operators are given by Eq. (\ref{zn_op}).

In the following we restrict ourselves for both models to the massless case, $m=0$, and the subspace of vanishing total charge, $\sum_n\left(\phi^\dagger_n\phi_n-\frac{1}{2}\left[\mathds{1}-(-\mathds{1})^n\right]\right)=0$, for which analytical results are available \cite{Schwinger1962}. No big qualitative changes are expected for the massive case.

\subsection{Numerical approach}
We study the model Hamiltonians using standard MPS techniques to compute the ground state and simulate the time evolution. The MPS ansatz for a system of $N$ sites with open boundary conditions is of the form
\begin{align*}
 |\Psi\rangle = \sum_{i_1,i_2,\dots,i_N} A_1^{i_1}A_2^{i_2}\dots A_N^{i_N}|i_1\rangle |i_2\rangle \dots |i_N\rangle,
\end{align*}
where $A_k^{i_k}$ are $D\times D$--dimensional complex matrices for $1<k<N$ and $A_1^{i_1}$ ($A_N^{i_N}$) is a row (column) vector. Each superscript $i_k$ ranges from $1$ to the dimension $d_k$ of the local Hilbert space of site $k$, and $|i_k\rangle_{k=1}^{d_k}$ forms a basis of the local Hilbert space. The number $D$, the bond dimension of the MPS, determines the number of variational parameters in the ansatz and limits the amount of entanglement which can be present in the state. For convenience in the simulations, we use an equivalent spin formulation of each Hamiltonian \cite{Banks1976}, which can be obtained via a Jordan-Wigner transformation on the fermionic degrees of freedom.

In our simulations, we are interested in different aspects. First, we would like to determine the effect of using finite-dimensional Hilbert spaces for the gauge degrees of freedom. To study this, we compute the ground state for each of the models by variationally minimizing the energy as described in Ref. \cite{Verstraete2004}. 
Second, to analyze the performance of the adiabatic preparation scheme, in particular, the effect of noise, we need to simulate time evolution. In order to compute the evolution numerically we split the Hamiltonian into two sums, each containing only mutually commuting three-body terms, and approximate the time evolution operator via a second-order time-dependent Suzuki-Trotter decomposition \cite{Suzuki1993}. This allows us to simulate the time evolution of the models with MPS \cite{Vidal2003,Verstraete2004a,Daley2004}, as long as the system stays close to the ground state \cite{Schuch2008} (a detailed review of MPS methods can be found in Refs. \cite{Schollwoeck2011,Verstraete2008}).

In our simulations, errors may originate from two main sources. Both in the ground state and in the dynamical simulations, the bond dimension employed is limited. Nevertheless, this source of error is controlled by choosing a sufficiently large $D$. In the dynamical simulations, an additional source of error arises from the Suzuki-Trotter decomposition of the time evolution operator. This error can be controlled via the time step size used for the splitting (a more detailed analysis of our numerical errors for the results presented in the following sections is reported in Appendix \ref{app:extrapolation}).

\section{\label{sec:finite_d}Effect of the finite dimension}
In order to analyze the effect of using finite-dimensional systems to represent the gauge degrees of freedom, we study the ground states of the truncated cQED and $\Zn$ models for different (odd) values of $d$, ranging from $3$ to $9$, and compare them to the case of the lattice Schwinger model.

In a lattice calculation, in which the goal is to extract the continuum limit, simulations need to be run at different values of the lattice spacing. Hence, we have also explored the effect of the finite $d$ for various lattice spacings, $ga$, and for several system sizes. As a figure of merit, we analyze the ground-state energy density, $\omega=E_0/2Nx$, and compare the values in the thermodynamic limit obtained in each case to those from finite-size extrapolations of the lattice Schwinger model. In the previous expression $N$ is the number of fermionic sites in the chain, $x$ is related to the lattice spacing as $x=1/(ga)^2$, and $E_0$ denotes the ground-state energy of the dimensionless Hamiltonian $2H/ag^2$ [see Appendix \ref{app:scaling}, Eq. (\ref{dimless_hamiltonian}), for the explicit expression] \footnote{The quantities $\omega$, $E_0$, and $x$ are frequently used in lattice calculations for the Schwinger model \cite{Banks1976,Hamer1982,Banuls2013} and we adapt to this convention for better comparability}. To get the energy density in the thermodynamic limit, we first compute the ground-state energy, $E_0$, for each set of parameters $(N,d,x)$ for various bond dimensions $D$, which allows us to extrapolate $D\to\infty$ and estimate our numerical errors. Subsequently, we extrapolate $N\to\infty$ for each combination of $(x,d)$ which yields the values for $\omega$ in the thermodynamic limit (details about the extrapolation to the thermodynamic limit can be found in Appendix \ref{app:extrapolation}).

\begin{figure}[htp!]
\centering
\includegraphics[width=0.5\textwidth]{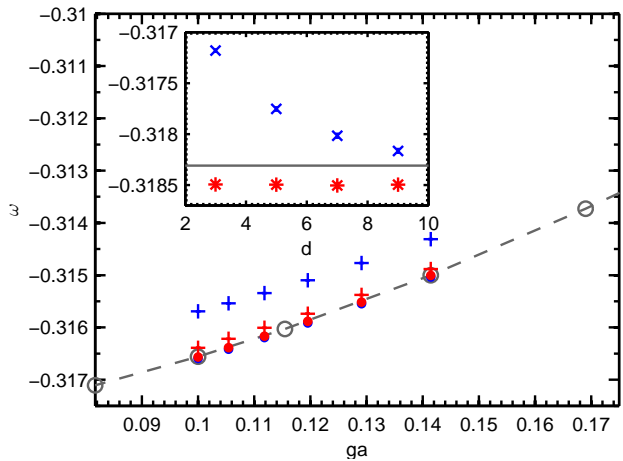}
\caption{Thermodynamic limit for the energy density for various values of $x$. Crosses show the values for the truncated cQED model for $d=3$ [upper (blue) crosses] and $d=9$ [lower (red)  crosses]. Circles show the values for the $\Zn$ model for $d=3$ (blue) and $d=9$ (red), which are almost identical. Values obtained for the Schwinger model are shown in gray. Inset: Values obtained by extrapolating $x\to\infty$ for the continuum energy density for the truncated cQED model [(blue) \X's] and the $\Zn$ model [(red) asterisks]. The horizontal gray line represents the value for the Schwinger model in the massless case, $-1/\pi$. In both cases the error bars from the extrapolation procedure are smaller than the markers.}
\label{fig:E_vs_N0}
\end{figure}

In our simulations we explore system sizes such that $N$ ranges from $50$ to $200$, and lattice spacings corresponding to values of $x\in[50,100]$. Our results are shown in Fig. \ref{fig:E_vs_N0} \footnote{Here we show the energy density, as this quantity allows an extrapolation to the thermodynamic and to the continuum limit for the range of parameters studied. We observe that also other quantities, such as the chiral condensate, approach the values of the Schwinger model with increasing Hilbert space dimension of the links. However, the extrapolation process for the condensate is a lot more delicate and it is not expected to yield very accurate results in the parameter regime we have explored, even for the full model \cite{Banuls2013a}.}. We observe that the truncated cQED model converges to the values of the Schwinger model with increasing value of $d$. By contrast, the $\Zn$ model already yields very accurate results even for low values of $d$ and the level of accuracy stays practically constant for larger $d$.

In our range for $x$, we can also attempt a continuum limit extrapolation for each set of values (see inset in Fig. \ref{fig:E_vs_N0}) \cite{Crewther1980,Durr2005}. Here we observe that the truncated cQED model approaches rapidly the exact value for increasing $d$, whereas for the $\Zn$ model the continuum extrapolation is already quite close to it for $d=3$ and there is almost no change for larger $d$. This is consistent with our observations for the thermodynamic limit, where the results in the $\Zn$ case are already very accurate for each lattice spacing, even for small $d$. However, one should take into account that the values of $x$ used in this work are relatively small to extrapolate to the continuum \cite{Banuls2013}, which is likely the source of larger systematic errors not taken into account here (a more detailed description of the extrapolation procedure and error estimation is given in Appendix \ref{app:extrapolation}). Hence the level of error due to the finite-dimensional Hilbert spaces is expected to be already smaller than that of the extrapolation.

\section{\label{sec:evo}Adiabatic preparation of the ground state}
Given a physical system which effectively implements one of these models, the nontrivial vacuum state could, in principle, be constructed using an adiabatic step \cite{Georgescu2014}. In this step one starts with an initial state, which is the ground state of a simpler Hamiltonian and easy to prepare. Subsequently the interactions are then slowly switched on to reach the desired model.

For both models considered here, a valid initial state could be the strong coupling ground state ($x=0$) in the \emph{physical} (i.e. Gauss law fulfilling) subspace, which is a simple product state with the odd (even) sites occupied (empty) and the links carrying no flux, $|\psi_0\rangle=\su \fz \sd \fz \su \fz \sd \dots$ \cite{Banks1976,Hamer1982}. In the previous expression, the bold numbers represent the occupation of the sites.
The coupling strength can be tuned by changing $x$ from $0$ to $x_\mathrm{F}$. Provided the change is slow enough, the adiabatic theorem ensures that the final state will be close to the ground state for $x_\mathrm{F}$.

\begin{figure}[htp!]
\centering
\includegraphics[width=0.5\textwidth]{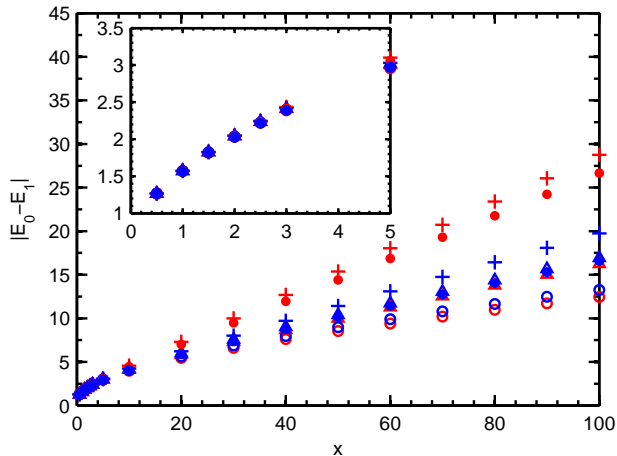}
\caption{Gap between the ground state and the first excited state in the Gauss law fulfilling sector for the $\Zn$ model and the truncated cQED model. Open symbols represent the values for the $\Zn$ model for $N=50$  (triangles) and $N=100$ (circles). Values for the truncated cQED model are represented by the crosses ($N=50$) and circles ($N=100$). Red markers indicate $d=3$; blue markers, $d=9$. Inset: The region for small values of $x$ shown in greater detail. All data points were computed with $D=60$.}
\label{fig:gap}
\end{figure}

The resources required to successfully perform this preparation are dominated by the total time $T$ needed for an adiabatic enough evolution, which depends on the inverse gap of the Hamiltonian. As our model Hamiltonians are of the from (\ref{hamiltonian}), it can be directly seen that the gap vanishes in the massless case for $x=0$. For finite values of $x$, Fig. \ref{fig:gap} reveals that the gap starts to grow with increasing $x$, and the growth in the region of small $x$ is almost independent of system size $N$ and Hilbert space dimension $d$ for both models. Thus the change of the Hamiltonian at early times (or \emph{while $x$ is small}) has to be very slow, whereas it is rather insensitive after reaching larger values of $x$.

To analyze the performance of a quantum simulation that runs this adiabatic preparation, we simulate a ramping of the parameter $x$ form $0$ to a value of $x_\mathrm{F}=100$ which corresponds to the smallest lattice spacing used in the previous section. We use a function $x(t) = x_\mathrm{F}\cdot(t/T)^3$ that turns out to be flat enough at the beginning in our evolution simulations. 

In order to probe the scaling of the required time with system size and other parameters, we deem an evolution successful only if the overlap with the exact ground state is above a minimum value ($0.99$). We monitor the overlap between the evolved state and the exact ground state for various values of $t$, where the exact ground state is computed using the method from the previous section \footnote{To compute the ground state variationally, we use a significantly higher bond dimension of $D=100$ than for the evolution to make sure we have a quasi-exact state.}.

As the cQED ($\Zn$) Hamiltonian commutes with $G_n^{\mathrm{cQED}}$ ($U_n^{\Zn}$) independently from the value of $x$, and our initial state is in the physical subspace, the Gauss law will be fulfilled at any time during the preparation procedure. As a consistency check for the numerics, nevertheless, we monitor whether the simulated state stays in the physical subspace with a total charge equal to 0, which is characterized by $U_n^{\Zn}=\mathds{1}$ ($G_n^\mathrm{cQED}=0$) for the $\Zn$ (truncated cQED) model. Therefore a violation results in a finite expectation value of the observable $P^{\Zn}=\sum_n\left(U_n^{\Zn}-\mathds{1}\right)^\dagger \left(U_n^{\Zn}-\mathds{1}\right)$ ($P^\mathrm{cQED}=\sum_n G_n^{\mathrm{cQED} \dagger} G_n^\mathrm{cQED}$) in the $\Zn$ (truncated cQED) case that can be detected during the evolution.
\begin{figure}[htp!]
\centering
\includegraphics[width=0.5\textwidth]{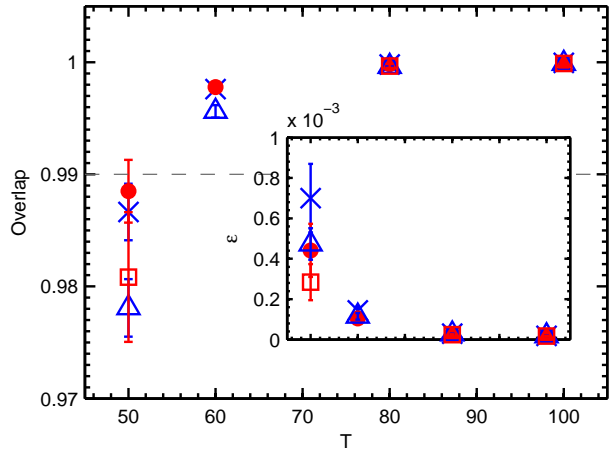}
\caption{Truncated cQED model with $D=50$. Final overlap with the exact ground state at the end of the adiabatic preparation as a function of the total evolution time. The (blue) \X's represent the data for $N=50$, $d=3$; (blue) triangles for $N=100$, $d=3$; (red) circles, for $N=50$, $d=9$; and (red) squares for $N=100$, $d=9$. Error bars were obtained from the difference in results with bond dimension $D=50$ vs $D=30$. Inset: Relative error of the energy with respect to the exact ground state.}
\label{fig:cQED_evo}
\end{figure}
\begin{figure}[htp!]
\centering
\includegraphics[width=0.5\textwidth]{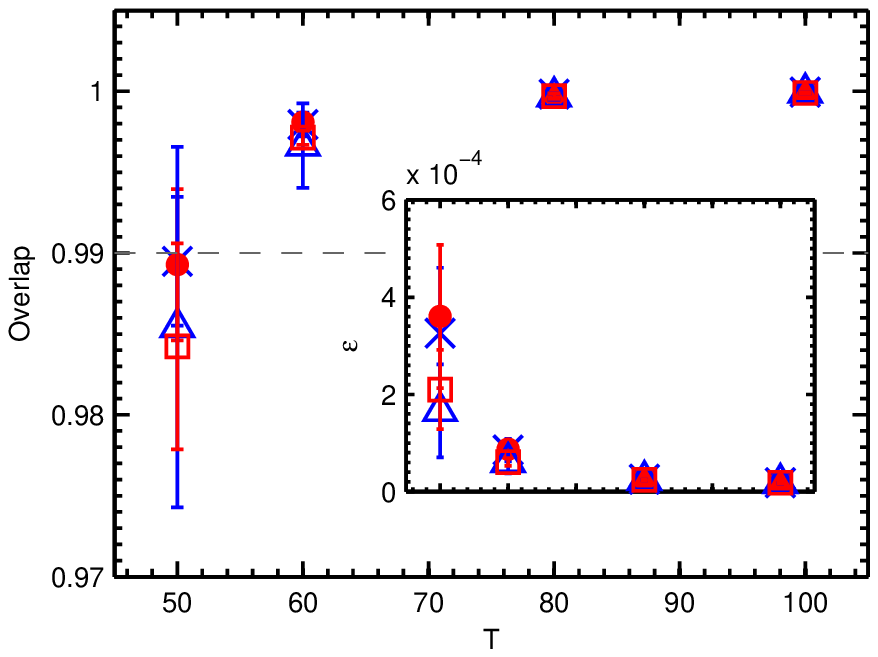}
\caption{$\Zn$ model with $D=50$. Final overlap with the exact ground state at the end of the adiabatic preparation as a function of the total evolution time. The (blue) \X's represent the data for $N=50$, $d=3$; (blue) triangles, for $N=100$, $d=3$; (red) circles for $N=50$, $d=9$; (red) squares for $N=100$, $d=9$. Error bars were obtained from the difference in results with bond dimension $D=50$ vs $D=30$. Inset: Relative error of the energy with respect to the value for the exact ground state.}
\label{fig:Zn_evo}
\end{figure}

The results obtained by the preparation procedure described above are shown in Figs. \ref{fig:cQED_evo} and \ref{fig:Zn_evo}. For all the data presented here, we see that the expectation values of $P^\nu$ during the evolution indeed stay down to 0 up to machine accuracy, where $\nu=\mathrm{cQED},\Zn$ labels the appropriate model. We find that for the chosen parameters we can obtain overlaps higher than $0.99$ for both models around a total evolution time of $T=60$ and the results still improve until $T=80$, where we reached an overlap close to 1 and the error bars are already smaller than the markers. The relative error $\varepsilon$ in the energy with respect to the exact ground-state energy (see insets in Figs. \ref{fig:cQED_evo} and \ref{fig:Zn_evo}) shows a similar behavior. Remarkably, for the range of parameters we have studied, the results are almost independent of the system size, $N$, and the Hilbert space dimension, $d$, as can be checked in Figs. \ref{fig:cQED_evo} and \ref{fig:Zn_evo}, where data are shown for $N=50$ and $100$. This is in accordance with our observation that the gap does not depend on the system size and the Hilbert space dimension for small values of $x$ (see Fig. \ref{fig:gap}).

\section{\label{sec:noisy_evo}Effect of broken gauge invariance}
One crucial question for the quantum simulation of LGT is whether the nonfundamental character of the gauge invariance will limit the power of the method. Even though it has been shown  that it is possible to have models where the invariance is ensured at the level of interactions among the quantum systems \cite{Zohar2013,Zohar2013a,Banerjee2013}, external sources of noise that do not fulfill the gauge symmetry will likely be present in an experiment.

\begin{figure}
\centering
\includegraphics[width=0.5\textwidth]{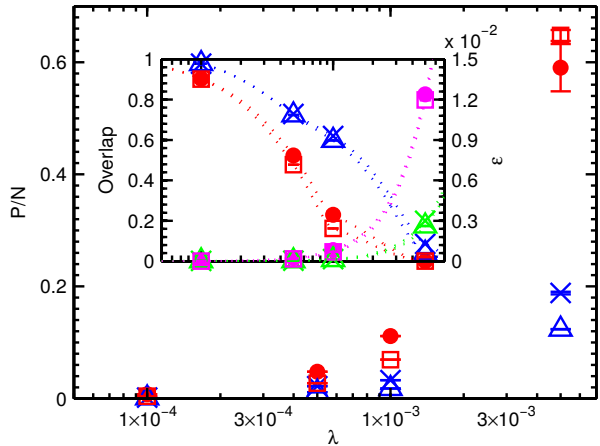}
\caption{Truncated cQED model. Penalty energy per site at the end of the noisy adiabatic preparation as a function of the noise strength. The [blue (green)] \X's represent the values for $N=50$, $d=3$; the [blue (green)] triangles, the $N=100$, $d=3$ case; the [red (magenta)] circles, the $N=50$, $d=5$ case; and the [red (magenta)] squares, the $N=100$, $d=5$ case. Error bars were computed the same way as in the noiseless case. Inset: Overlap (blue and red symbols) and relative error in energy (green and magenta symbols) with respect to the noise-free exact ground state. As a guide for the eye, data points are connected.}
\label{fig:cQED_Overlap_vs_noise}
\end{figure}

\begin{figure}
\centering
\includegraphics[width=0.5\textwidth]{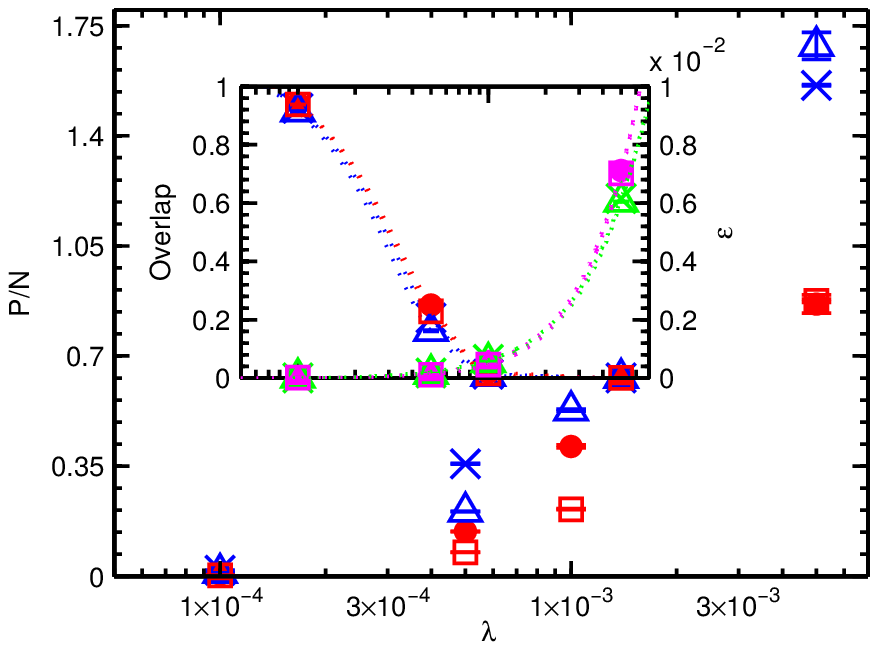}
\caption{$\Zn$ model. Penalty energy per site at the end of the noisy adiabatic preparation as a function of the noise strength. The [blue (green)] \X's represent the values for $N=50$, $d=3$; the [blue (green)] triangles, the $N=100$, $d=3$ case; the [red (magenta)] circles, the $N=50$, $d=5$ case; and the [red (magenta)] squares, the $N=100$, $d=5$ case. Error bars were computed the same way as in the noiseless case. Inset: Overlap (blue and red symbols) and relative error in energy (green and magenta symbols) with respect to the noise-free exact ground state. As a guide for the eye, data points are connected.}
\label{fig:Zn_Overlap_vs_noise}
\end{figure}

In order to study the effect of such nongauge symmetric contributions, we add a noise term to the Hamiltonian, which is given by $\sum_n\lambda x(t)(L^+_n+L^-_n)$ for the $\Zn$ case and by $\sum_n\lambda x(t)(a^\dagger_nb_n+b^\dagger_na_n)$ for the truncated cQED case. This could represent some noise that occurs in the experimental setup implementing the interactions and is, thus, proportional to their strengths, $x$. The parameter $\lambda$ would then be the relative strength of the noise. We simulate the same adiabatic protocol as in the previous section, for a total time $T=100$, which ensures success of the evolution as described earlier, under different levels of noise and for the same values of the other parameters $(N,d,x_\mathrm{F})$ studied before. In addition to the overlap with respect to the exact ground state, we quantify the violation of the Gauss law per particle $P^\nu/N$ for each case. The results are shown in Figs. \ref{fig:cQED_Overlap_vs_noise} and \ref{fig:Zn_Overlap_vs_noise}.

We observe that even small levels of noise ($\lambda=5\times 10^{-4}$) result in finite values of $P^\nu/N$ and a drastic reduction of the final overlap with the ground state. Nevertheless, the relative error in the energy stays below $2\%$ for both models. Consequently, if the noise can be controlled relative to the value of $x$, the predictions for some ground-state observables may still be quite accurate although the gauge invariance is broken.

Figures \ref{fig:cQED_Overlap_vs_noise} and \ref{fig:Zn_Overlap_vs_noise} also show that the quantity $P^\nu/N$ does not show a strong dependence on the system size. To get an estimation of the scaling, we computed perturbatively the first nonvanishing contribution to the expectation value of $P^\nu$ \cite{Franson2002}. We found that $P^\nu\propto N,\lambda^2$ for fixed $T,x_\mathrm{F}$ independently of the ramping. Specifically for our choice of $x(t)$, we find $P^\nu$ to be proportional to $(\lambda x_\mathrm{F})^2t^8 N$ (see Appendix \ref{app:scaling}). Consequently, independent of the system size, $P^\nu/N$ is proportional to $\lambda^2$ for a fixed value of $t$, consistent with our data.

\section{\label{sec:conclusion}Conclusion}
Using MPS techniques, we have studied numerically two particular proposals for quantum simulation of the lattice Schwinger model. These methods allow us to address three important questions that affect the feasibility of quantum simulation for more general LGT.

First, we have shown that although the finite dimension of the physical systems that represent gauge variables on the links may affect the ground state of the model, the results converge rapidly as this dimension is increased.  In particular, for the truncated cQED model, we observed fast convergence to the exact ground state of the Schwinger model for $d$ ranging from $3$ to $9$. For the $\Zn$ model, the results with $d=3$ are already extremely close to those of the full model.

Second, we have discussed an adiabatic preparation protocol for the ground state starting from a simple product state. Our results suggest that the preparation is feasible and that the initial part of the evolution is crucial for its success. With a suitable choice of $x(t)$, we can obtain an overlap of more than $0.99$ with the exact ground state for both models. Most remarkably, the required total time (for a given final value $x_\mathrm{F}$) is practically insensitive to the system size and the physical dimension of the gauge variables, in accordance with the observed gap.

Finally we have shown that the procedure for adiabatic preparation of the ground state is to some extent robust to noninvariant terms as the energy can still be reliably determined up to a certain noise level. This is promising, as it demonstrates that even if the gauge invariance is broken, which could happen due to noise or at the fundamental level of interactions among the basic ingredients, the proposals do not immediately lose their predictive power. Furthermore, the scaling of our results is in good agreement with a perturbative calculation.

In our study, we have proposed a polynomial ramp for $x$, slow enough to achieve the desired preparation. However, with the observation that the gap opens with increasing values of $x$ and the results from the perturbative calculation, one could think about designing an optimized ramp $x(t)$. Furthermore, optimal control concepts could also be helpful to design optimized ramps \cite{Walmsley2003}. On the one hand, this could allow shorter total evolution times while keeping the same level of overlap with the exact ground state in the noise-free case. On the other hand, one could possibly achieve a better scaling of the Gauss law violation with time in the presence of noninvariant terms and therefore improve the robustness of the preparation scheme proposed.

\section*{Acknowledgements}
We thank Benni Reznik and Erez Zohar for helpful discussions. This work was partially funded by the EU through SIQS Grant No. FP7 600645.

\bibliographystyle{h-physrev}
\bibliography{Papers_MPQ}

\appendix
\section{\label{app:extrapolation}Numerical errors}
In this Appendix we provide details about the extrapolation procedure and our estimation of errors. For the ground-state calculations we run the variational ground-state search \cite{Verstraete2004} for different system sizes, $N$, different lattice spacings, $x$, and several (odd) physical dimensions, $d$, of the link variables, ranging from $3$ to $9$. For each combination $(N,d,x)$ we increase the bond dimension until the ground-state energy converges up to a predefined relative accuracy. For the truncated cQED case we find $D=100$ together with relative accuracy $10^{-6}$ to be sufficient for all the studied parameters, while for the $\Zn$ model we go up to $D=200$ and a relative accuracy of $10^{-12}$. Our final energy value is extrapolated linearly in $1/D$  using the two largest computed bond dimensions and the error is estimated as the difference from the largest $D$ result.

For the results shown in Fig. \ref{fig:E_vs_N0}, we perform a finite-size extrapolation for each pair $(x,d)$, using the same functional form as in \cite{Banuls2013}:
\begin{align*}
\frac{E_0}{2Nx} = \omega+\frac{c_1}{N}+\mathcal{O}(N^{-2}).
\end{align*}
Similarly to the procedure described in \cite{Banuls2013}, we extrapolate to the continuum from each set of values for a given $d$, by fitting the ground-state energy densities obtained in the previous step to a quadratic function in $1/\sqrt{x}=ga$. This limit is expected to be only of limited precision since the values used in this paper, $x\in[50,100]$, are still far away from the continuum, which constitutes a source of error much more important than that of the particular fit.

In the case of time evolution we have an additional source of error due to the second-order time-dependent Suzuki-Trotter approximation \cite{Suzuki1993} of the time evolution operator. For the results presented in Secs. \ref{sec:evo} and \ref{sec:noisy_evo} we have tried different time steps and a value of $\Delta t=0.001$ turns out to be sufficiently small, so that the errors are much below the observed effects. The large error bars in Figs. \ref{fig:cQED_Overlap_vs_noise} and \ref{fig:Zn_Overlap_vs_noise} for small $T$ are due to the limited bond dimension. In these cases the evolution is not adiabatic enough to stay close to the ground state and one ends up in a superposition state which cannot be well approximated by a MPS with our values of $D=30,50$. As one can see, for longer total evolution times, where one stays close to the ground state, this effect vanishes and the simulations converge with a small $D$.

\section{\label{app:scaling}Analytic estimation of the effect of gauge invariance breaking perturbations}
To get an idea how the violation of the Gauss law scales in the case of noisy evolution, we compute the lowest order contribution to $\langle \psi(t)|P^\nu|\psi(t)\rangle$ using perturbation theory following Ref. \cite{Franson2002}. For clarity we simply write $P$ and suppress, for the rest of this section, the index labeling the model. Additionally, to keep the equations short, we introduce $U_n$, which refers to $U^{\Zn}_n-\mathds{1}$ ($G^\mathrm{cQED}_n$) in the $\Zn$ (truncated cQED) case.

Starting from a dimensionless version of our model Hamiltonian, $W=2H/ag^2 $, we use an equivalent spin formulation,
\begin{align}
\begin{aligned}
W(t)=&\sum_{n=1}^{N-1}\left(L^z_n\right)^2+ \frac{\mu}{2}\sum_{n=1}^N(-1)^n\left(\sigma^z_n+\mathds{1}\right)\\
&+x(t)\sum_{n=1}^{N-1}\left(\sigma^+_nL^+_n\sigma^-_{n+1} + \mathrm{h.c.}\right),
\end{aligned}
\label{dimless_hamiltonian}
\end{align}
where $\mu=2m/ag^2$, and add the noise term $\sum_n\lambda x(t)\left(\bar{L}^+_n + \bar{L}^-_n\right)$ to it,
\begin{align*}
 \tilde{W}(t)=W(t)+\sum_n \lambda x(t)\left(\bar{L}^+_n + \bar{L}^-_n\right),
\end{align*}
where $\bar{L}^\pm_n$ refers to $L^\pm_n$ for the $\Zn$ model and to $a^\dagger_nb_n$ ($b^\dagger_na_n$) for the truncated cQED model, and therefore coincides with the $L^\pm_n$ operators for this model up to a constant. For small times $t$ and small values of $\lambda$ we can treat the noise term as a perturbation to the Hamiltonian $W(t)$. The contributions to $\langle \psi(t)|P|\psi(t)\rangle$ are given by subsequent commutators of $P$ with the Hamiltonian
\begin{align}
\begin{aligned}
 &\langle \psi(t)|P|\psi(t)\rangle = \langle \psi_0|P|\psi_0\rangle\\
  &+\frac{1}{i}\int_0^t\mathrm{d}t'\langle \psi_0|[P,\tilde{W}(t')]|\psi_0\rangle\\
  &+\frac{1}{i^2}\int_0^t\mathrm{d}t'\int_0^{t'}\mathrm{d}t''\langle \psi_0|\left[[P,\tilde{W}(t')],\tilde{W}(t'')\right]|\psi_0\rangle\\
  &+\dots ,
\end{aligned}
\label{perturbation_theory}
\end{align}
where $\psi_0$ is the initial state; in our case this is a product state fulfilling the Gauss law. As the unperturbed Hamiltonian commutes with $U_n$ and $\bar{L}^\pm_n|\psi_0\rangle$ is still an eigenstate of $U_n$ which is orthogonal to $|\psi_0\rangle$, it is immediately clear that the first contribution occurs at second order and the double commutator reduces to

\begin{alignat*}{3}
 &\Bigl[[P\bigr.,\tilde{W}(t')]\left.,\tilde{W}(t'')\right] = && &&\\ 
&=-\lambda^2x(t')x(t'')\sum_{n,m,k}&&\Bigl(&&\langle\psi_0|\bar{L}^+_mU^\dagger_n U_n \bar{L}^-_k|\psi_0\rangle \\
& &&+&&\langle\psi_0|\bar{L}^-_m U^\dagger_nU_n\bar{L}^+_k|\psi_0\rangle\\
& &&+&&\langle\psi_0|\bar{L}^+_k U^\dagger_nU_n\bar{L}^-_m|\psi_0\rangle\\
& &&+&&\langle\psi_0|\bar{L}^-_k U^\dagger_nU_n\bar{L}^+_m|\psi_0\rangle\Bigr)\\
&=-2\lambda^2x(t')x(t'')\sum_{n}&&\Bigl( &&\langle\psi_0|\bar{L}^-_nU^\dagger_n U_n \bar{L}^+_n|\psi_0\rangle \\
& &&+&&\langle\psi_0|\bar{L}^+_n U^\dagger_nU_n\bar{L}^-_n|\psi_0\rangle \Bigr).
\end{alignat*}

In the second step we have used that $\bar{L}^\pm_m|\psi_0\rangle$ are eigenstates of $U_n$, with nonzero eigenvalue iff $m=n$, and that $\langle\psi_0|\bar{L}^\mp_k\bar{L}^\pm_m|\psi_0\rangle=c^\pm_m\cdot\delta_{k,m}$ with a constant $c^\pm_m$. Thus there are only contributions if $n=k=m$ and we are left with a single sum. The two different matrix elements appearing in the sum are simply giving two constants, hence the sum can be estimated as $cN$ with a constant $c$. Plugging this back into Eq. (\ref{perturbation_theory}), we obtain
\begin{align*}
 \langle \psi(t)|P|\psi(t)\rangle \approx 2\lambda^2cN\int_0^t\mathrm{d}t'x(t')\int_0^{t'}\mathrm{d}t''x(t'').
\end{align*}

For our $x(t)=x_\mathrm{F}\cdot(t/T)^3$ the integrals can be easily solved yielding
\begin{align}
 \langle \psi(t)|P|\psi(t)\rangle \approx 2(\lambda x_\mathrm{F})^2\frac{t^8}{32T^6}cN.
 \label{penalty_scaling}
\end{align}

\begin{figure}[H]
\centering
\includegraphics[width=0.47\textwidth]{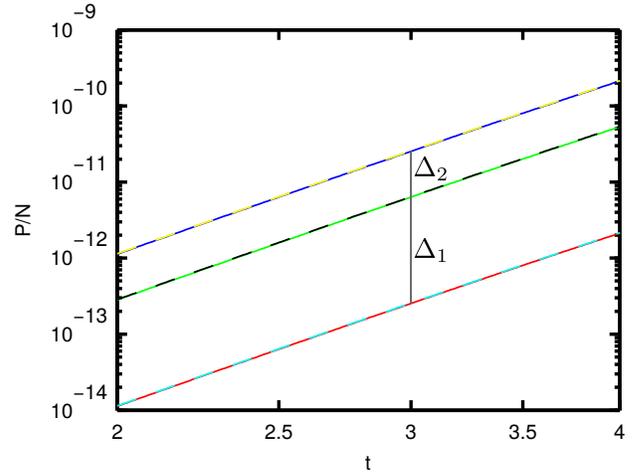}
\caption{Truncated cQED model. Penalty energy per site as a function of time for $d=3$ (both axes are on a logarithmic scale). The vertical gray line indicates the point in time where we determined the offsets $\Delta_1$ and $\Delta_2$. The lower [red ($N=50$) and cyan ($N=100$)] dashed lines show the values for $\lambda=1\times 10^{-4}$, the middle [green ($N=50$) and black ($N=100$)] dashed lines show the values for $\lambda=5\times 10^{-4}$, and the upper [blue ($N=50$) and yellow ($N=100$)] dashed lines show the values for $\lambda=1\times10^{-3}$.}
\label{fig:cQED_Penalty_scaling_d3}
\end{figure}

\begin{figure}[H]
\centering
\includegraphics[width=0.47\textwidth]{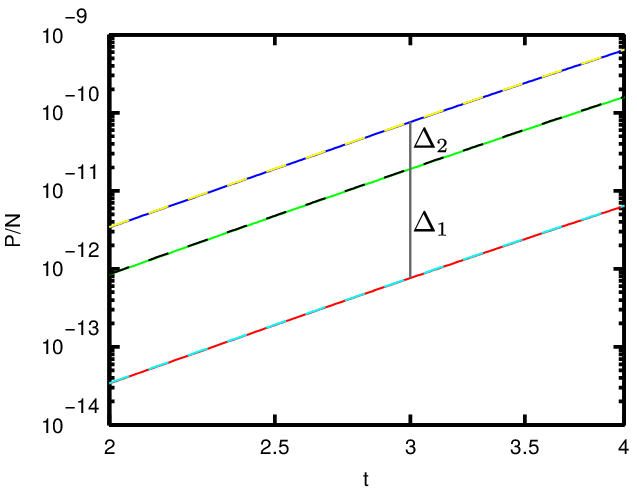}
\caption{Truncated cQED model. Penalty energy per site as a function of time for $d=5$ (both axes are on a logarithmic scale). The vertical gray line indicates the point in time where we determined the offsets $\Delta_1$ and $\Delta_2$. The lower [red ($N=50$) and cyan ($N=100$)] dashed lines show the values for $\lambda=1\times10^{-4}$, the middle [green ($N=50$) and black ($N=100$)] dashed lines show the values for $\lambda=5\times10^{-4}$, and the upper [blue ($N=50$) and yellow ($N=100$)] dashed lines show the values for $\lambda=1\times10^{-3}$.}
\label{fig:cQED_Penalty_scaling_d5}
\end{figure}

\begin{figure}[H]
\centering
\includegraphics[width=0.47\textwidth]{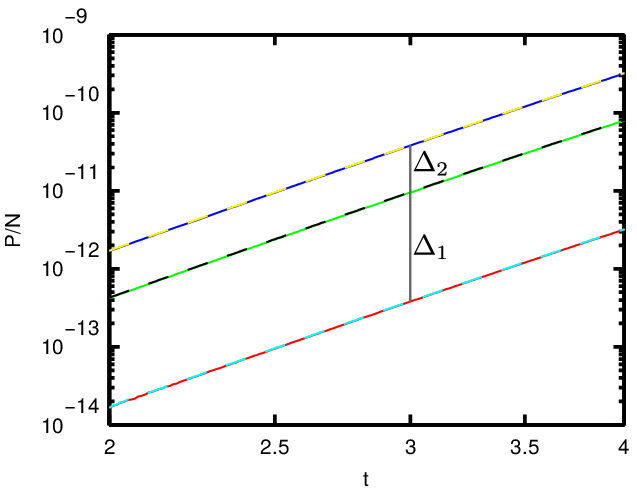}
\caption{$\Zn$ model. Penalty energy per site as a function of time for $d=3$ (both axes are on a logarithmic scale). The vertical gray line indicates the point in time where we determined the offsets $\Delta_1$ and $\Delta_2$. The lower [red ($N=50$) and cyan ($N=100$)] dashed lines show the values for $\lambda=1\times10^{-4}$, the middle [green ($N=50$) and black ($N=100$)] dashed lines show the values for $\lambda=5\times10^{-4}$, and the upper [blue ($N=50$) and yellow ($N=100$)] dashed lines show the values for $\lambda=1\times10^{-3}$.}
\label{fig:Zn_Penalty_scaling_d3}
\end{figure}

\begin{figure}[H]
\centering
\includegraphics[width=0.47\textwidth]{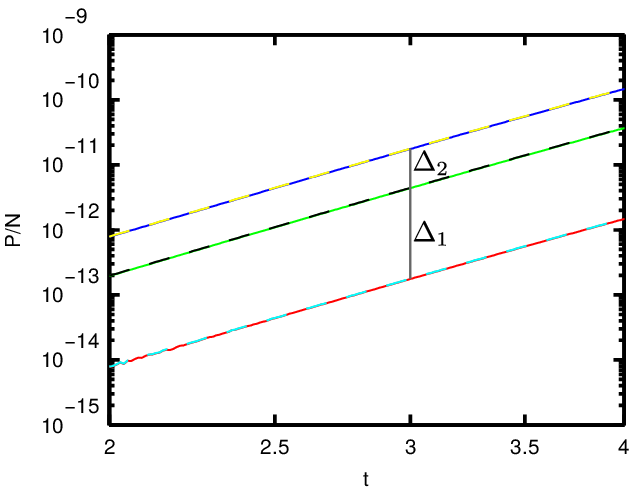}
\caption{$\Zn$ model. Penalty energy per site as a function of time for $d=5$ (both axes are on a logarithmic scale). The vertical gray line indicates the point in time where we determined the offsets $\Delta_1$ and $\Delta_2$. The lower [red ($N=50$) and cyan ($N=100$)] dashed lines show the values for $\lambda=1\times10^{-4}$, the middle [green ($N=50$) and black ($N=100$)] dashed lines show the values for $\lambda=5\times10^{-4}$, and the upper [blue ($N=50$) and yellow ($N=100$)] dashed lines show the values for $\lambda=1\times10^{-3}$.}
\label{fig:Zn_Penalty_scaling_d5}
\end{figure}

To numerically check this behavior, we plot $P/N$ as a function of time for both models (cf. Fig. \ref{fig:cQED_Penalty_scaling_d3}-\ref{fig:Zn_Penalty_scaling_d5}) for the three smallest values of noise used in Sec. \ref{sec:noisy_evo}. The time interval was chosen as close as possible to the beginning of the evolution but late enough to ensure that the values for $P/N$ are above the machine accuracy. These plots reveal that $P/N$ indeed shows a power law behavior in $t$ which is independent from $N$. 

To check the scaling with time, we can fit the data to extract the slope $m_\lambda$ for each case. This yields values between $7.5544$ and $7.5589$ for all cases presented in Fig. \ref{fig:cQED_Penalty_scaling_d3}-\ref{fig:Zn_Penalty_scaling_d5} which is in good agreement with our calculations. Furthermore we can check the scaling with $\lambda$. From Eq. (\ref{penalty_scaling}) we obtain for the offset $\Delta$ between two curves with different noise levels $\lambda_1$ and $\lambda_2$
\begin{align*}
 \Delta &=|\log_{10}(\lambda_1^2)-\log_{10}(\lambda_2^2)| \\
 &=2\cdot|\log_{10}(\lambda_1)-\log_{10}(\lambda_2)|.
\end{align*}
For the values of $\lambda$ used here ($1\times 10^{-4}$, $5\times 10^{-4}$ and $1\times 10^{-3}$) this yields $\Delta_1= 1.3979$ and $\Delta_2=0.6021$. The values extracted from our numerical data for both models with various $N$ and $d$ show a relative deviation of at most $10^{-4}$ from these predictions, which indicates that there is almost no dependency on system size and Hilbert space dimension, in excellent agreement with our theoretical calculation.


\end{document}